\documentstyle[preprint,aps]{revtex} 
\begin{document} 

\draft 

\preprint{UTPT-96-13} 

\title{Lagrangian Formulation of a Solution to the Cosmological Constant Problem}

\author{J. W. Moffat} 

\address{Department of Physics, University of Toronto,
Toronto, Ontario, Canada M5S 1A7} 

\date{\today}

\maketitle 

\begin{abstract}%
A covariant Lagrangian formulation of a solution to the cosmological constant
problem, based on vizualising the fluctuations of the vacuum energy as a
non-equilibrium process with stochastic behaviour, is presented. The variational
principle yields equations of motion for the cosmological ``constant" $\Lambda$,
treated as a dynamical field, together with an equation for a Lagrange multiplier
field $\phi$, and the standard Einstein field equations with a variable cosmological
constant term. A stochastic model of $\Lambda$ yields a natural explanation for the smallness or
zero value of the constant in the present epoch and its large value in an era of
inflation in the early universe.
\end{abstract} 

\pacs{ } 

A recent model for solving the cosmological constant problem has been 
proposed\cite{Moffat},
in which the vacuum energy is treated as a fluctuating environment with stochastic
behaviour. In the following, we shall present a covariant formulation based
on a Lagrangian density, which yields classical equations incorporating
Einstein's gravitational field equations, upon which a
stochastic treatment using a Wiener process can be developed. 

Although the model
uses methods of critical phenomena and non-equilibrium statistical mechanics to
model the vacuum energy, it can be considered as a phenomenological description
of the kinds of behaviour that could be expected in a more fundamental quantum
gravity theory. It is generally agreed that the cosmological constant problem cannot
be solved within the context of a purely classical theory of gravity. However, no
satisfactory quantum gravity theory has been formulated, so it is hoped that
our model can shed light on the solution to the problem without the full apparatus of
such a theory.

The Lagrangian density is given by
\begin{equation}
{\cal L}={\cal L}_R+{\cal L}_{\Lambda}+{\cal L}_M,
\end{equation}
where 
\begin{mathletters}
\begin{eqnarray}
{\cal L}_R&=&\sqrt{-g}g^{\mu\nu}R_{\mu\nu},\\
{\cal L}_{\Lambda}&=&-2\sqrt{-g}[\Lambda
+(\Lambda_{,\mu}u^\mu-\alpha\Lambda+\Lambda^2)\phi],
\end{eqnarray}
\end{mathletters}
and $\Lambda=\Lambda(x)$ is the variable cosmological ``constant",
treated as a dynamical field, $\phi$ is a Lagrange multiplier field,
$u^\mu=dx^\mu/d\tau$ is an observer's four-velocity along a world
line in spacetime, $\alpha$ is a constant and ${\cal L}_M$ is a matter Lagrangian
density.  A variation of  ${\cal L}$ with respect
to $\phi$ and $\Lambda$ yields the equations of motion:
\begin{mathletters}
\begin {eqnarray}
\label{Lambdaequation}
\Lambda_{,\mu}u^\mu-\alpha\Lambda+\Lambda^2&=&0,\\
\label{phiequation}
\frac{1}{\sqrt{-g}}(\sqrt{-g}u^\mu\phi)_{,\mu}+(\alpha-2\Lambda)\phi-1&=&0.
\end{eqnarray}
\end{mathletters}
Varying ${\cal L}$ with respect to $g^{\mu\nu}$ and using
(\ref{Lambdaequation}) gives
\begin{equation}
\label{Einsteinequation}
R_{\mu\nu}-\frac{1}{2}g_{\mu\nu}R+\Lambda g_{\mu\nu}=8\pi GT_{\mu\nu}.
\end{equation}
The kinematical variable $u^\mu$ is the four-velocity of a fluid
element along a world line in spacetime, associated with a fluid with density $\rho$ and 
pressure $p$, so we do not vary ${\cal L}_{\Lambda}$ with respect to $u^\mu$.

The cosmological constant enters through the vacuum energy density:
\begin{equation}
T_{V\mu\nu}=-\rho_Vg_{\mu\nu}=-\frac{\Lambda_V}{8\pi G}g_{\mu\nu}.
\end{equation}
Today, $\Lambda$ has the small value, $\Lambda 
< 10^{-46}\,\,\hbox{GeV}^4$,
whereas generic inflation models require that $\Lambda$ has a relatively large value
during the inflationary epoch. This is the source of the cosmological constant
problem.

The line element in the Friedmann-Robertson-Walker model is
\begin{equation}
d\tau^2=dt^2-R^2(t)\biggl\{\frac{dr^2}{1-kr^2}+r^2d\theta^2+r^2\sin^2\theta 
d\phi^2\biggr\},
\end{equation}
where $k=-1,0,+1$ and we have used comoving coordinates with $u^\mu=(0,0,0,1)$.
Then, Eqs.(\ref{Lambdaequation}), (\ref{phiequation}) and (\ref{Einsteinequation})
become
\begin{mathletters}
\begin{eqnarray}
\label{Lambda2}
{\dot{\Lambda}}=\alpha\Lambda-\Lambda^2,\\
\dot{\phi}+\frac{3{\dot{R}}}{R}\phi+(\alpha-2\Lambda)\phi-1=0,\\
H^2\equiv \biggl(\frac{{\dot{R}}}{R}\biggr)^2=\frac{8\pi G\rho_M}{3}
+\frac{\Lambda}{3}-\frac{k}{R^2},%
\end{eqnarray}
\end{mathletters}
where ${\dot\Lambda}=d\Lambda/dt$, $\rho_M$ denotes the mass density, $H$ is
the Hubble constant, and
in the following $H_0$ and $t_0$ denote the present values of $H$ and $t$,
respectively.
We define
\begin{equation}
\Omega_{\hbox{tot}}\equiv\Omega_M+\Omega_{\Lambda}=1-\Omega_k,
\end{equation}
where $\Omega=8\pi G\rho/3H^2$.

We shall treat the vacuum energy as a fluctuating environment and 
consider $\Lambda$ as a variable characterizing the state of this system. 
The parameter $\alpha$ in Eq.(\ref{Lambda2}) corresponds
to the difference between the growth and decline of particle-antiparticle annihilation
in the vacuum, while the second term is a self-restriction term which limits the
growth of $\Lambda$.

In ref.(1), we considered the situation in which the
vacuum fluctuations are rapid compared
with $\tau_{\hbox{macro}}=\alpha^{-1}$, which defines the macroscopic scale of
time evolution. We assumed that the parameter $\alpha$ can be written as
$\alpha_t=\alpha+\sigma\xi_t$, in which $\alpha$ is the average value,
$\xi_t$ is Gaussian noise and $\sigma$ measures the intensity of the vacuum
fluctuations. Let us write Eq.(\ref{Lambda2}) as 
\begin{equation}
d\Lambda_t=(\alpha\Lambda_t-\Lambda_t^2)dt+\sigma\Lambda_t dW_t
=f(\Lambda_t)+\sigma g(\Lambda_t)dW_t,
\end{equation}
where $dW_t$ is a Wiener process. The probability density $p(x,t)$, which describes
the $\Lambda$ distribution, satisfies the Fokker-Planck equation:
\begin{equation}
\label{Fokker}
\partial_t p(x,t)=-\partial_x[(\alpha x-x^2)p(x,t)]+\frac{\sigma^2}{2}
\partial_{xx}(x^2 p(x,t)).
\end{equation}
The diffusion process is restricted to the positive real half line and $0$ and
$\infty$ are intrinsic boundaries, because $g(0)=0$ and $f(\infty)=-\infty$. The
probability of the diffusion process reaching infinity as $t\rightarrow\infty$ is zero,
since infinity is a natural boundary. Moreover, zero is a natural boundary if $\alpha
>\sigma^2/2$, so neither boundary is accessible and no boundary
conditions need be imposed on the Fokker-Planck equation. For $\alpha <
\sigma^2/2$, it can be shown that zero is an {\it attracting} boundary.

The stationary-state solution for the probability density, $p_s(x)$, of 
Eq.(\ref{Fokker}) is given by\cite{Horsthemke}
\begin{equation}
p_s(x)=Nx^{(2\alpha/\sigma^2)-2}\exp\biggl(-\frac{2x}{\sigma^2}\biggr).
\end{equation}
The normalization constant $N$ is
\begin{equation}
N^{-1}=\biggl[\biggl(\frac{2}{\sigma^2}\biggr)^{2(\alpha/\sigma^2)-1}\biggr]^{-1}
\Gamma\biggl(\frac{2\alpha}{\sigma^2}-1\biggr),
\end{equation}
where $\Gamma$ denotes the $\Gamma$-function. If $p(x,t)$ is integrable between
$0$ and $\infty$, then a stationary state solution
exists when $\alpha> \sigma^2/2$. If it does not exist, then the probability density
will be concentrated at zero, i.e., $p(\Lambda) = \delta(\Lambda)$ for $\alpha <
\sigma^2/2$.

For $0 < \alpha < \sigma^2/2$, the vacuum fluctuations dominate over the growth
or decline of $\Lambda$, although the value zero is still the most probable value
for $\Lambda$, since the distribution function has a vertical slope at
$\Lambda=0$. Because we are using a continuous variable, $\Lambda$ never
reaches the boundary zero in a finite time.

When $\alpha > \sigma^2/2$, the growth of
$\Lambda$ dominates the influence of the vacuum fluctuations, and in the
neighborhood of zero the probability of $\Lambda=0$ drops to zero. For the
stochastic model there are two transition points described by different order
parameters. At $\alpha=\sigma^2/2$ real growth of $\Lambda$ becomes possible
corresponding
to a change from a degenerate random variable for steady-state behavior to
a stochastic variable; the boundary at $\alpha=0$ switches from attracting to
natural. Secondly, there is the transition point $\alpha=\sigma^2$ which
corresponds to a qualitative change in the stochastic variable $\Lambda$
with no change in the nature of the boundary. The probability of $\Lambda=0$
drops abruptly to zero.

The following scenario can be deduced from our model. In the inflation
era, the intensity of vacuum fluctuations is large and $\alpha > \sigma^2$,
causing a second-order phase transition and a maximum in $\Lambda$ not
near zero. This corresponds to the large vacuum energy needed to drive
inflation\cite{Linde}. As the universe expands the intensity of vacuum fluctuations
decreases
and for $0<\alpha<\sigma^2/2$ or $\sigma^2/2<\alpha<\sigma^2$ the probability
density is largest when $\Lambda$ is non-vanishing and small, which can lead to a
current value of
$\Lambda_0$ that can be used to fit the observational data. 
If the stationary probability density $p_s$ does not exist for $\alpha<\sigma^2/2$,
then $\Lambda=0$
is a stationary point; the drift and diffusion vanish simultaneously for $\Lambda=0$
and $p(\Lambda)=\delta(\Lambda)$. This corresponds to the case when 
$\Lambda$ is vanishingly small.

Thus, our model provides a natural explanation, in terms of non-equilibrium
stochastic processes in an expanding universe, for the behavior of $\Lambda$
required to fit observational data and still be consistent with inflationary models.

Ongoing searches\cite{Perlmutter} for Type Ia supernovae show that 
$\Omega_{\Lambda} < 0.47$ (at 95\% confidence level for spatially flat
$\Lambda$ models). Moreover, for $\Lambda\not=0$ a larger fraction
of QSOs would be gravitationally lensed and QSO surveys give 
$\Omega_{\Lambda}\leq 0.7$\cite{Kochanek}. Cold dark matter models (CDM)
for large scale stucture formation, which include a cosmological constant, yield
a better fit to the shape of the observed spectrum of galaxy clustering than
does the standard $\Omega_M=1$ CDM models, using 
$h\equiv H_0/(100 {\rm km/sec/Mpc})=0.7$,
$\Omega_{\Lambda}=0.6$ and a baryon density with $\Omega_B=0.0255$,
consistent with primordial nucleosynthesis\cite{Peacock}.  However, the amplitude
for the $\Lambda$ CDM models is too high compared to the data, a problem that
persists at all scales.

The problem of the age of the universe is also alleviated in $\Lambda$ models. An
analysis of the cosmological data showed that for $\Omega_{\Lambda}=0.65\pm 0.1,
\Omega_M=1-\Omega_{\Lambda}$ and a small tilt: $0.8<n<1.2$, models exist
which are consistent with the available data and an inflationary spatially flat
universe\cite{Steinhardt}.

Models based on $\Lambda$ treated as a scalar dynamical field\cite{Moffat2} have
been found
to partially resolve observational problems. The observations of gravitationally
lensed QSOs yield a less restrictive upper bound on $H_0t_0$ in these 
models\cite{Dodelson}. They may also provide a solution to the size of the
amplitude problem, since although the shape of the spectrum is the same
as that of the $\Lambda$ CDM model with $\Omega_{\Lambda}=0.6$, the
dynamical $\Lambda$ model yields a lower amplitude and therefore gives
a better fit to the galaxy clustering data\cite{Dodelson}.

\acknowledgments
I thank M. A. Clayton for helpful discussions. This work was supported by the Natural Sciences
and Engineering Research
Council of Canada.

\end{document}